\documentclass[journal,twoside,web]{ieeecolor}
\usepackage{generic}
\usepackage{cite}
\usepackage{amsmath,amssymb,amsfonts}
\usepackage{algorithmic}
\usepackage{graphicx}
\usepackage{algorithm,algorithmic}
\usepackage{hyperref}
\hypersetup{hidelinks=true}
\usepackage{textcomp}
\usepackage{multirow}
\usepackage{caption}
\def\BibTeX{{\rm B\kern-.05em{\sc i\kern-.025em b}\kern-.08em
    T\kern-.1667em\lower.7ex\hbox{E}\kern-.125emX}}
\begin{document}
\title{A region and category confidence-based multi-task network for carotid ultrasound image segmentation and classification}
\author{Haitao Gan, Ran Zhou, Yanghan Ou, Furong Wang, Xinyao Cheng, Aaron Fenster, \IEEEmembership{Fellow, IEEE}
\thanks{This work has been submitted to the IEEE for possible publication. Copyright may be transferred without notice, after which this version may no longer be accessible. This work was supported by the National Natural Science Foundation of China under grant No. 62201203, the Natural Science Foundation of Hubei Province under grant No. 2021CFB282, the High-level Talents Fund of Hubei University of Technology under grant No. GCRC2020016. (Corresponding author: Ran Zhou.)}
\thanks{Haitao Gan is with the School of Computer Science, the Hubei University of Technology, Wuhan, Hubei, 430068, China (e-mail: htgan01@hbut.edu.cn). }
\thanks{Ran Zhou is with the School of Computer Science, the Hubei University of Technology, Wuhan, Hubei, 430068, China (e-mail: ranzhou@hbut.edu.cn). }
\thanks{Yanghan Ou is with the School of Computer Science, the Hubei University of Technology, Wuhan, Hubei, 430068, China (e-mail: 724334042@qq.com). }
\thanks{Furong Wang is with the Liyuan Hospital, Tongji Medical College, Huazhong University of Science and Technology, Wuhan, Hubei, 430074, China (e-mail: 386600121@qq.com).}
\thanks{Xinyao Cheng is with the Department of Cardiology, Zhongnan Hospital, Wuhan University, Wuhan, Hubei, 430068, China (e-mail: xycheng@whu.edu.cn). }
\thanks{Aaron Fenster is with the Imaging Research Laboratories, Robarts Research Institute, Western University, London,	ON N6A 5B7, Canada (e-mail: afenster@robarts.ca).}}

\maketitle

\begin{abstract}
The segmentation and classification of carotid plaques in ultrasound images play important roles in the treatment of atherosclerosis and assessment for the risk of stroke. Although deep learning methods have been used for carotid plaque segmentation and classification, two-stage methods will increase the complexity of the overall analysis and the existing multi-task methods ignored the relationship between the segmentation and classification. These will lead to suboptimal performance as valuable information might not be fully leveraged across all tasks. Therefore, we propose a multi-task learning framework (RCCM-Net) for ultrasound carotid plaque segmentation and classification, which utilizes a region confidence module (RCM) and a sample category confidence module (CCM) to exploit the correlation between these two tasks. The RCM provides knowledge from the probability of plaque regions to the classification task, while the CCM is designed to learn the categorical sample weight for the segmentation task. A total of 1270 2D ultrasound images of carotid plaques were collected from Zhongnan Hospital (Wuhan, China) for our experiments. The results showed that the proposed method can improve both segmentation and classification performance compared to existing single-task networks (i.e., SegNet, Deeplabv3+, UNet++, EfficientNet, Res2Net, RepVGG, DPN) and multi-task algorithms (i.e., HRNet, MTANet), with an accuracy of 85.82\% for classification and a Dice-similarity-coefficient of 84.92\% for segmentation. In the ablation study, the results demonstrated that both the designed RCM and CCM were beneficial in improving the network's performance. Therefore, we believe that the proposed method could be useful for carotid plaque analysis in clinical trials and practice.
\end{abstract}


\section{Introduction}
\label{sec:introduction}
\IEEEPARstart{C}{ardiovascular} disease (CVD) is the global leading cause of death and disability, and ischemic stroke is one of the primary causes of death among CVD patients \cite{vos2020global}. Atherosclerotic plaques can rupture, resulting in thrombus formation and vascular stenosis that can lead to corresponding hemodynamic changes and ischemic cardiovascular and neurovascular events \cite{gisteraa2017immunology}. Because the carotid arteries (CA) have a simple geometry and are easily accessible, they are commonly used to visualize and assess carotid plaques and their potential risk for causing a stroke. Therefore, identifying carotid plaques and assessing the risk they pose to the patient is a critical aspect of treating atherosclerosis and reducing the risk of stroke \cite{hogberg2014carotid}. 

Carotid ultrasound is widely used in carotid plaque identification and assessment due to its non-invasive and low-cost characteristics. Over the past decade, researchers have developed various ultrasound biomarkers for the quantification of carotid plaques and showed them to be useful for monitoring changes in the carotid arteries, including intima-media thickness (IMT) \cite{bots1997common}, total plaque area (TPA) \cite{spence2002carotid}, and total plaque volume (TPV) \cite{wannarong2013progression}. Several studies have demonstrated correlations between plaque echogenicity and its vulnerability to rupture. For example, hyperechoic plaques are often associated with calcification, which have traditionally been regarded as stable atheromas \cite{aburahma2002carotid}. Hypoechoic plaques are always considered to be at a higher risk for the development of a stroke \cite{polak1998hypoechoic}. Mixed-echoic plaques are heterogeneous and composed of a mixture of hypoechoic, isoechoic, and hyperechoic components, which are also related to the risk of stroke. Thus, different types of carotid plaques pose different risks of causing cerebrovascular events such as stroke. Generally, the appearance of carotid plaques in B-mode ultrasound images can be divided into three types: hyperechoic plaque, hypoechoic plaque, and mixed-echoic plaque \cite{aburahma2002carotid,kobayashi2000detection}. 

Measurement of these ultrasound-based biomarkers requires segmentation of plaque boundaries and identification of the carotid plaque type using classification methods. Thus, segmentation and classification of carotid plaques in ultrasound images are two critical tools in carotid ultrasound image analysis. Many computer-aided diagnostic algorithms have been proposed for carotid plaque segmentation and classification, which can be divided into two categories: traditional image processing-based methods or deep learning-based approaches. 

Traditional segmentation methods include level sets, active contour models, snake models, Gaussian mixture models, and geometric priors, which have alleviated the manual segmentation process \cite{destrempes2011segmentation,loizou2014integrated,cheng2013fully}. However, these methods are sensitive to contour initialization and image quality, resulting in segmentation inaccuracy and instability that cannot meet the demands of clinical applications. As a result, deep learning-based methods have become the mainstream research direction for carotid plaque segmentation, with most studies focusing on deep learning network structures, loss function design, and post-processing of the segmentation results. Jain et al. developed hybrid deep learning segmentation models (i.e., UNet, UNet+, SegNet, SegNet-UNet, and SegNet-UNet+) for atherosclerotic plaques in the internal carotid artery using B-mode ultrasound images \cite{jain2021hybrid}. They further used seven U-series architectures for measuring the area of the plaque far-wall of the common carotid (CCA) and internal carotid arteries (ICA) in B-mode ultrasound images \cite{jain2022far}. Mi et al. proposed an MBFF-Net for carotid plaque segmentation in ultrasound images by designing a multi-branch feature fusion module to extract multiple scales and different contexts and exploited the prior information of the carotid artery wall \cite{mi2021mbff}. Li et al. proposed a U-shaped CSWin transformer for carotid artery segmentation in 3D ultrasound images, where hierarchical CSWT modules were used to capture the rich global context information in the 3D image \cite{lin2023method}. We also used a U-Net to generate TPA from B-model carotid ultrasound images \cite{zhou2021deepJBHI} and further proposed ensemble learning approaches to reduce segmentation inconsistency, smooth segmentation contours, and improve the accuracy and robustness of plaque segmentation models \cite{zhou2023adaptively}.

Many other works have focused on ultrasound carotid plaque classification. Christodoulou et al. extracted ten different texture feature sets and used the statistical k-nearest neighbor method for atherosclerotic carotid plaque classification \cite{christodoulou2003texture}. Kyriacou et al. used two classifiers, the Probabilistic Neural Network and a Support Vector Machine (SVM), to classify atherosclerotic carotid plaques into symptomatic or asymptomatic types \cite{kyriacou2009classification}. Prahl et al. proposed a percentage white feature to classify the echogenicity in carotid plaques \cite{prahl2010percentage}. Acharya et al. extracted several grayscale features and input them into an SVM classifier for plaque tissue and classification \cite{acharya2012plaque}. Engelen et al. used 3D texture features to predict vascular events from 3D carotid plaque ultrasound images \cite{van2014three}. However, these traditional plaque classification methods rely on feature extraction algorithms that cannot accurately and comprehensively extract carotid plaque features. However, in recent years, researchers have successfully applied deep learning methods to carotid ultrasound image classification. Lekadir et al. used a convolutional neural network (CNN) to identify plaque compositions from carotid ultrasound images \cite{lekadir2016convolutional} and Saba et al. implemented six deep artificial intelligence models for carotid ultrasound plaque characterization using images obtained from a multicenter study \cite{saba2021multicenter}.

While deep learning techniques have found utility in carotid plaque assessment, it is important to note that these methods have typically concentrated on either segmentation or classification tasks independently. Consequently, a common practice in these approaches involves the adoption of a two-stage process for carotid plaque image analysis. This two-stage process can lead to increased complexity in the overall analysis, and the effectiveness of the classification performance is sensitive to the accuracy of the prior segmentation task. Therefore, we believe that the implementation of a multi-task deep learning framework can be advantageous for both segmentation and classification tasks. Performing segmentation and classification together allows them to inform each other, leading to more accurate and robust results. Additionally, this approach has the potential to simplify the processing stream and reduce computation time. Zhu et al. designed a multi-task UNet for saliency prediction and disease classification on chest X-ray Images \cite{zhu2022jointly}. Wang et al. developed a multi-task synergetic network to segment and classify polyps by using a multi-scale training strategy \cite{wang2023efficient}. Ling et al. proposed a multi-task attention network by using the reverse addition attention to fuse areas in multi-level layers of UNet for medical image segmentation and classification \cite{ling2023mtanet}. 

However, the existing multi-task methods may not effectively integrate knowledge or features learned from one task into another. This can result in suboptimal performance as valuable information might not be fully leveraged across all tasks. Furthermore, the application of a multi-task network for segmentation and classification of carotid plaques has not yet been developed.

In this study, we present a region and category confidence-based multi-task network (RCCM-Net) for carotid ultrasound image segmentation and classification aimed at effectively exploring the relationship between segmentation and classification to facilitate both tasks. For segmentation, we have designed a sample category confidence module (CCM) to generate a category weight for each sample, which encourages the network to focus on hard samples. For the classification task, we propose a region confidence module (RCM) to obtain segmented probability maps from multiple levels of the segmentation network. We then utilize these probability maps to fuse features in the classification task, allowing the classification task to focus on high-confidence plaque regions. To our knowledge, this is the first work using multi-task learning for carotid ultrasound image analysis. The main contributions of this study are as follows:

\begin{figure*}[!htbp]
	\begin{center}
		\includegraphics[width=\textwidth,height=\textheight,keepaspectratio]{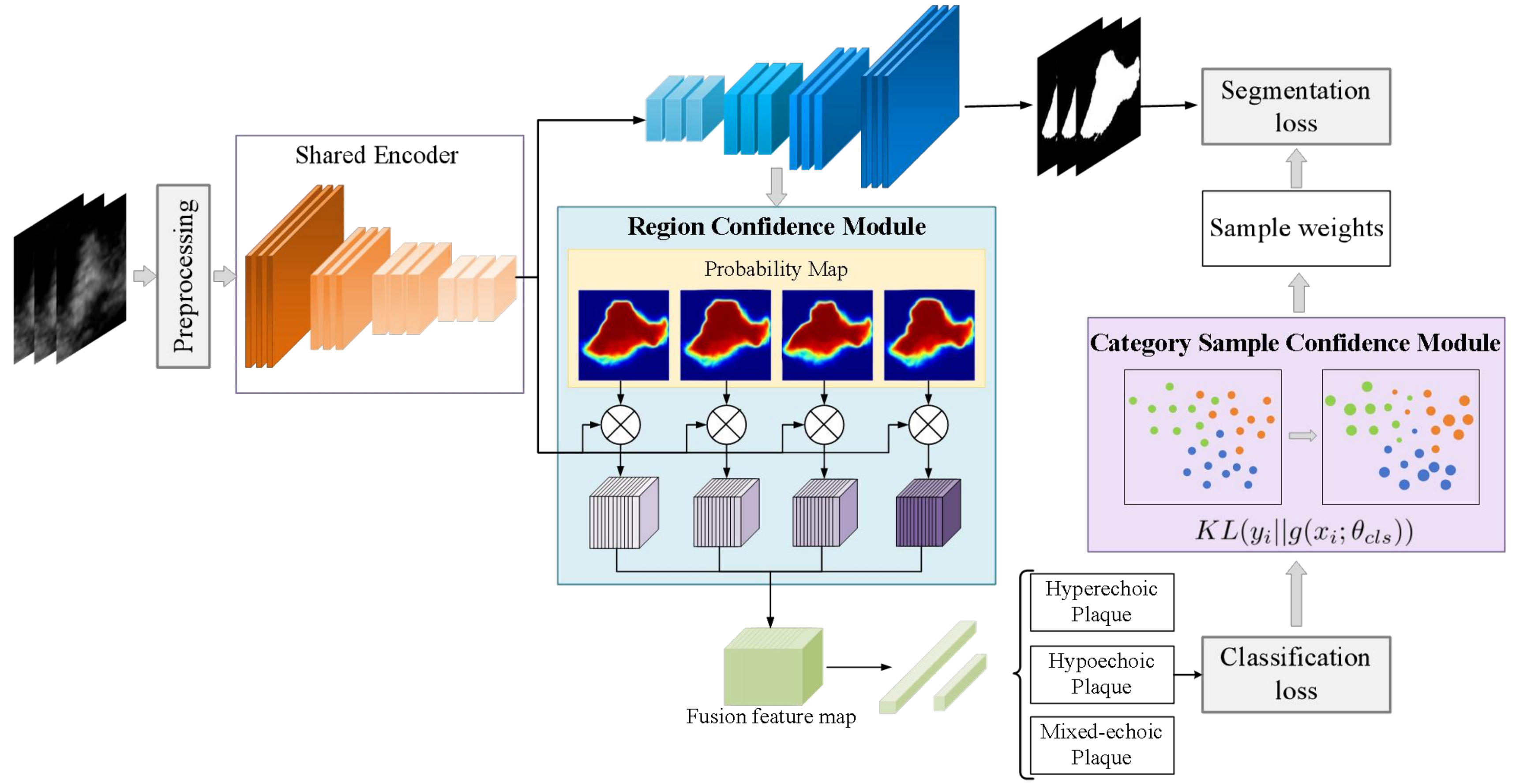}
	\end{center}
	\caption{Framework of RCCM-Net including a segmentation task (upper branch) and a classification task (lower branch), where the two branches share the encoder (original blocks). The region confidence module (RCM) uses the segmentation probability to generate region confidence maps as region-weights to integrate the feature maps in the classification task. The category sample confidence module (CCM) generates a category confidence level for each sample (image) to cause the network to focus on hard samples in the segmentation loss.}
	\label{fig:framework}
\end{figure*}

(1) This study introduces a novel multi-task learning framework (RCCM-Net) that addresses both ultrasound plaque segmentation and classification tasks. This framework enables a single network to effectively handle these two related tasks simultaneously with high accuracy and low variability, making it possibly suitable for clinical use.

(2) The category confidence module (CCM) is proposed to assign category confidence weights for each sample in segmentation loss to direct the network focus toward hard sample learning. These category confidence weights are generated by calculating the difference between the probability distribution of the classification results and the probability distribution of the labels. 

(3) The region confidence module (RCM) is developed to enable the network to learn features from high-confidence plaque regions. RCM combines the outputs of different levels in the segmentation decoder to generate region weights and merge the feature maps obtained in the feature extraction layers of the classification network.
 
\section{Method}
The proposed multi-task learning framework leverages the complementary information of carotid plaque segmentation and classification tasks to enhance the overall performance of the system. Figure \ref{fig:framework} shows the general framework of RCCM-Net, which contains a segmentation task branch and a classification task branch. By segmenting the carotid plaque, the network learns crucial information about the plaque size and shape, enabling the classification branch to concentrate on features in plaque regions while minimizing the influence of irrelevant background information. The classification task provides valuable category information about plaques and integrates this category information into the loss function of the segmentation to improve the segmentation performance of difficult plaque images. The network incorporates two new modules: the region confidence module (RCM) and the category sample confidence module (CCM), which mutually enhance the performance of the classification and segmentation tasks. RCM generates region confidence maps by using the pixel-wise probability from different level outputs of the segmentation task. These region confidence maps are utilized as region-weights to integrate the feature maps in the classification task. CCM generates a category confidence level of each sample (image) by comparing the predicted and labeled probability distributions of the classification task. These category confidence levels are applied as sample weights to make the network focus on hard samples in the segmentation loss. The segmentation and classification tasks share the feature extraction encoder during training.

\subsection{Region Confidence Module (RCM)}
\begin{figure}[!htbp]
	\begin{center}
		\includegraphics[width=0.5\textwidth,height=\textheight,keepaspectratio]{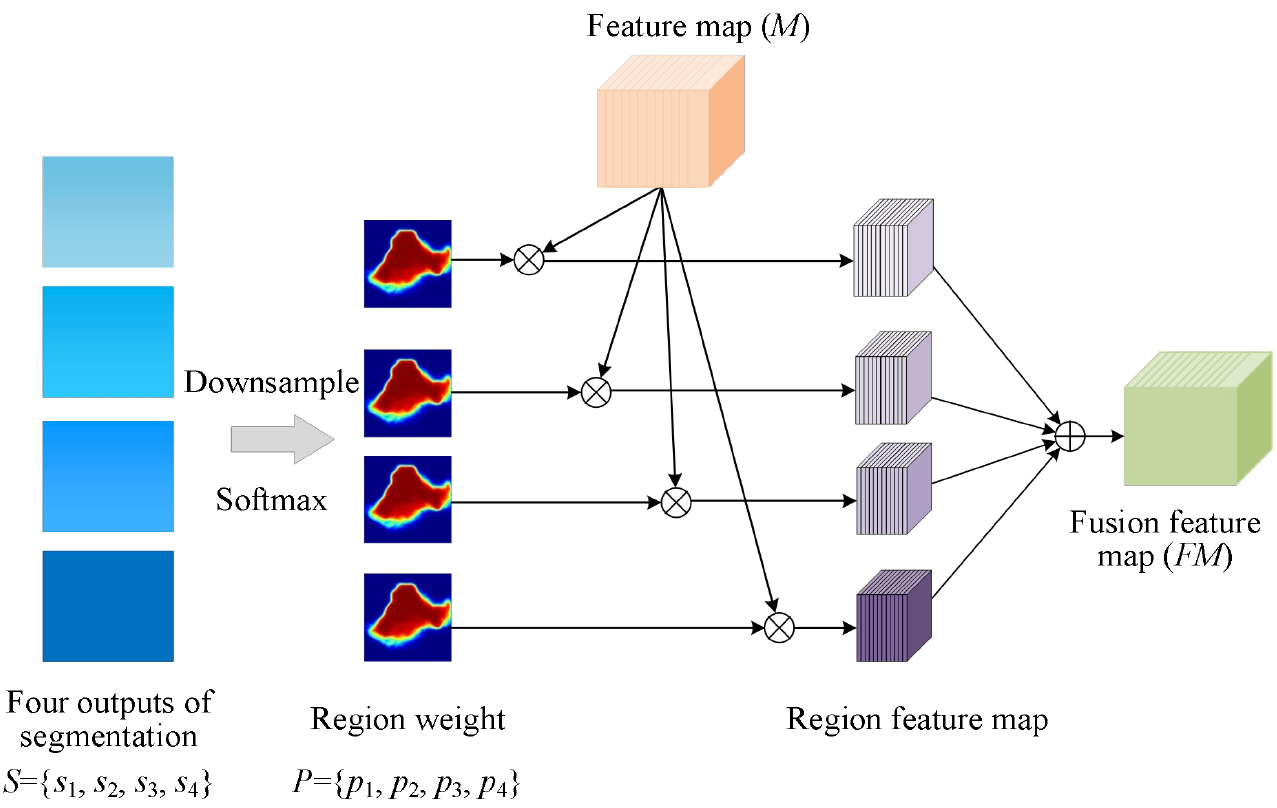}
	\end{center}
	\caption{The details of the RCM. The four different level outputs of the segmentation decoder are used to obtain four region-weight maps for the region feature fusion task.}
	\label{fig:RCM}
\end{figure}

To facilitate the classification task, RCM combines the different level outputs of the segmentation decoder to generate the region weights. These weights are then used to fuse the feature maps obtained in the feature extraction layers (encoder). Figure \ref{fig:RCM} shows the detailed architecture of RCM. First, the four outputs are downsampled to the same size as the feature maps obtained in the encoder. After that, a Softmax function is applied to obtain the probability maps. These probability maps obtained from the segmentation task are used as region-weights in the feature fusion step to improve the classification feature extraction. Each probability map is multiplied by the corresponding position of each channel of the feature map to obtain the region feature map. Finally, the four region feature maps are summed to obtain the fused feature map.

Assuming that the four outputs from different levels of the segmentation decoder are $S$=$\{s_1,s_2,s_3,s_4\}$, the region probability maps ($P$=$\{p_1,p_2,p_3,p_4\}$) are be obtained by 
\begin{equation}
	p_i=\frac{e^{F_{down}(s_i)}}{\sum_{i=1}^{4}{e^{F_{down}(s_i)}}}
\end{equation}
                                                    
where $F_{down}$ is a down-sampling function to make $S$ the same size as the feature maps ($M$) obtained in the encoder. The fusion feature maps ($FM$) are then formulated as
\begin{equation}
	FM=\sum_{i=1}^{4}{\alpha_ip_iM}
\end{equation}                                                   
where $\alpha_i$ are the hyperparameters. 

\subsection{Sample-Weight Module (CCM)}
\begin{figure}[!htbp]
	\begin{center}
		\includegraphics[width=0.5\textwidth,height=\textheight,keepaspectratio]{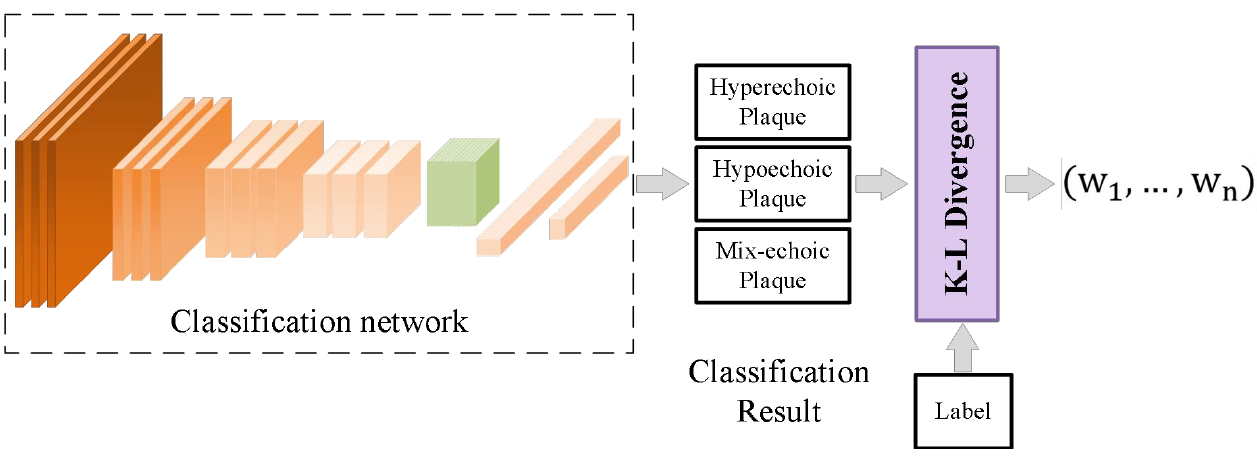}
	\end{center}
	\caption{The details of the CCM. K-L Divergence is used to calculate the confidence of the plaque-type prediction for each sample.}
	\label{fig:CCM}
\end{figure}

CCM uses the results of the classification task to assign a confidence level to each sample, which is then used as the sample-weight in the loss function of the segmentation task. Kullback-Leibler (K-L) Divergence is used to determine the confidence level by calculating the difference between the probability distribution of the classification results and the probability distribution of the labels. Finally, the result of K-L Divergence is used to represent the weights of each sample. Figure \ref{fig:CCM} shows the implementation details of CCM.
Assuming that $y_i$ denotes the label probability distribution of the $i$th input sample for classification and $g(x_i;\theta_{cls})$ denotes the probability distribution predicted by the classification network, the confidence level (sample-weight) for the $i$th sample is obtained by

\begin{equation}
	\omega_i=KL(y_i||g(x_i;\theta_{cls}))=\sum_{j=1}^{d}{y_{i,j}log(\frac{y_{i,j}}{g_j(x_i;\theta_{cls})})}
\end{equation}
where $d$ is the number of plaque classes.
For training samples that are incorrectly predicted by the encoder in CCM, the larger the result of K-L Divergence, the smaller the weight of the training samples. This is because a large K-L Divergence indicates that the predicted probability distribution is very different from the ground truth distribution, which indicates that the model is less confident in its prediction and assigns a smaller weight to the sample. But, if the K-L Divergence is small, it indicates that the predicted distribution is close to the ground truth distribution, and the model is more confident in its prediction, which results in a larger weight for the sample in the loss function.

\subsection{Loss Function}
Consider a carotid ultrasound dataset \begin{math}X=\{(x_i,y_i,z_i)|i\in[1,n], x_i\in \mathbb{R}^{W\times H}, y_i\in\mathbb{R}^{3\times1}, z_i\in\mathbb{R}^{2\times W\times H}\}\end{math}, where $x_i$ denotes the $i$th image, $y_i$ is the class label of the $i$th image, $z_i$ is the segmentation label of the $i$th image, and $y_i$ and $z_i$ are converted into one-hot formats to allow improved prediction accuracy. The segmentation loss function consists of the weighted cross-entropy loss ($L_{WCE}$) and the entropy loss ($L_{Entropy}$). 
\begin{equation}
L_{seg}=L_{WCE}(w_i,x_i,z_i;\theta_{seg})+L_{Entropy}(x_i;\theta_{seg})    
\end{equation}

where $w_i$ are obtained from the sample-weight module (CCM) and $\theta_{seg}$ is the set of segmentation network parameters.
The classification loss function consists of the general cross-entropy loss ($L_{CE}$) and entropy loss ($L_{Entropy}$). 
\begin{equation}
L_{cls}=L_{CE}(x_i,y_i;\theta_{cls})+L_{Entropy}(x_i;\theta_{cls})
\end{equation}
where $\theta_{cls}$ is the set of classification network parameters.
The entropy loss can promote the network to generate a one-hot distribution with a high probability for a single class, instead of using a flat distribution. Consequently, the network will predict results with increased certainty. Through the minimization of the entropy loss, the network can acquire the ability to produce more precise and confident predictions for each input sample.
Finally, the loss function of our multi-task framework is
\begin{equation}
Loss=L_{seg}+\lambda L_{cls}
\end{equation}
where $\lambda$ is a hyperparameter, and we set $\lambda=1$ in the experiments.

\subsection{Network Architecture}
\begin{figure}[!htbp]
	\begin{center}
		\includegraphics[width=0.5\textwidth,height=\textheight,keepaspectratio]{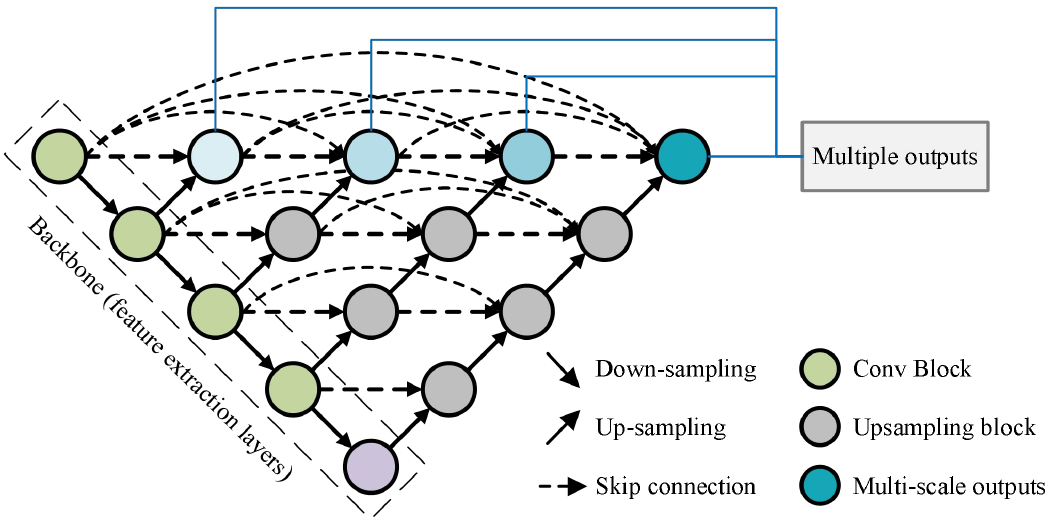}
	\end{center}
	\caption{Architecture of the UNet++ network for the segmentation task. It contains five U-Nets with varying depths that partially share the encoder (backbone), whereas the decoders are densely connected at the same resolution via redesigned skip connections.}
	\label{fig:architecture}
\end{figure}

U-Net-based network architectures have been widely used in medical image segmentation, and UNet++ is an improvement of U-Net and superior to other U-Net variants \cite{zhou2019unet++}. Figure \ref{fig:architecture} shows the architecture of UNet++, which fuses five convolutional blocks to extract multi-level features. Correspondingly, each convolutional block has a decoding path, except for the first convolutional block. Each decode path uses the upsampling block to convert feature maps of different sizes to the input size and to predict the segmentation mask of the input image. This architecture not only uses skip connections to connect feature maps from the encoder to each decoder path, but also uses skip connections to connect different decoder paths. The architecture enables flexible feature fusion, which helps to alleviate information loss when extracting features and converting images. 

In our multi-task algorithm, Residual blocks were implemented as the encoder in UNet++, which are used instead of the convolutional blocks in the backbones of UNet++. The Residual block uses two 1$\times$1 convolution kernels to change the number of channels in the feature map, a 3$\times$3 convolution kernel to extract features, and a short connection to sum the input and output of the convolution kernel \cite{he2016deep}. This block alleviates the degradation of the deep neural network, reduces the number of model parameters, and makes the deep neural network easy to train. 

\subsection{Evaluation Metrics}
To evaluate the performance of our proposed method comprehensively, we employed a set of widely used metrics to measure both segmentation and classification performance. Specifically, the Dice similarity coefficient (DSC), absolute plaque area difference ($|\Delta$PA$|$), Hausdorff distance (HD), average symmetric surface distance (ASSD), and Pearson correlation coefficient (PCC) were used to quantify the accuracy of our segmentation results. In addition, we used accuracy (ACC), precision, F1-score, and the Kappa coefficient to assess the classification performance of our algorithm.

DSC was used to calculate the similarity between the algorithm segmentation (A) and the manual segmentation (M), as
\begin{equation}
	DSC(A, M)=\frac{2|A\cap M|}{(|A|+|M|)}\times100\%
\end{equation} 

$|\Delta$PA$|$ was used to calculate the plaque area difference between the algorithm segmentation and manual segmentation as 
\begin{equation}
	|\Delta PA|= |PA_{alg}-PA_{man}|
\end{equation}
where $PA_{alg}$ is the algorithm-generated plaque area and $PA_{man}$ is the manually segmented plaque area. 

ASSD and HD were used to evaluate the distance between the algorithm and manual segmentation contours and are given by:
\begin{equation}
	\begin{split}
		ASSD(A,M)=&\frac{1}{2}(\frac{1}{|\partial R_A|}\sum_{p\in\partial R_A}d(p,\partial R_M)+\\
		&\frac{1}{|\partial R_M|}\sum_{p\in\partial R_M}d(p,\partial R_A)) 
	\end{split}
\end{equation}
\begin{equation}
	\begin{split}
		HD(A,M)=\max(
		& \max_{p\in\partial R_A} d(p,\partial R_M),\\ &\max_{p\in\partial R_M} d(p,\partial R_A))
	\end{split}
\end{equation}
where $\partial R_A$ is the algorithm segmentation surface and $d(p, \partial R_M)$  is the shortest Euclidean distance from a point $p$ to the manual segmentation surface $\partial R_M$; $d(p,\partial R_A)$  is defined in the same manner. 
The Pearson correlation coefficient (PCC) was used to measure the correlation between the algorithm and manually generated plaque areas.

In classification tasks, in addition to ACC, we also used precision and F1-score to measure the classification performance.
\begin{equation}
	F1-score=\frac{2\times Precision \times Recall}{Precision + Recall}
\end{equation}

\section{Experiments and Results}
\subsection{Data Acquisition}
A total of 1270 longitudinal carotid ultrasound images were obtained from Zhongnan Hospital of Wuhan University. The Zhongnan Hospital Institutional Review Board approved the use of the ultrasound data and all patients provided consent. These patients, who had risk factors such as hypertension or hyperlipidemia or had a history of vascular events, underwent ultrasound imaging of their carotid arteries (common, internal, and external). All images were acquired using an Acuson SC2000 (Siemens, Erlangen, Germany) ultrasound system with a 4-9 MHz linear array probe (9L4). Two experienced radiologists (Co-author X. Cheng and F. Wang) annotated the plaque contours and classified plaques into three categories (i.e., hyperechoic plaque, hypoechoic plaque, and mix-echoic plaque) according to the European Carotid Plaque Study Group criteria.

\begin{figure}[!htbp]
	\begin{center}
		\includegraphics[width=0.5\textwidth,height=\textheight,keepaspectratio]{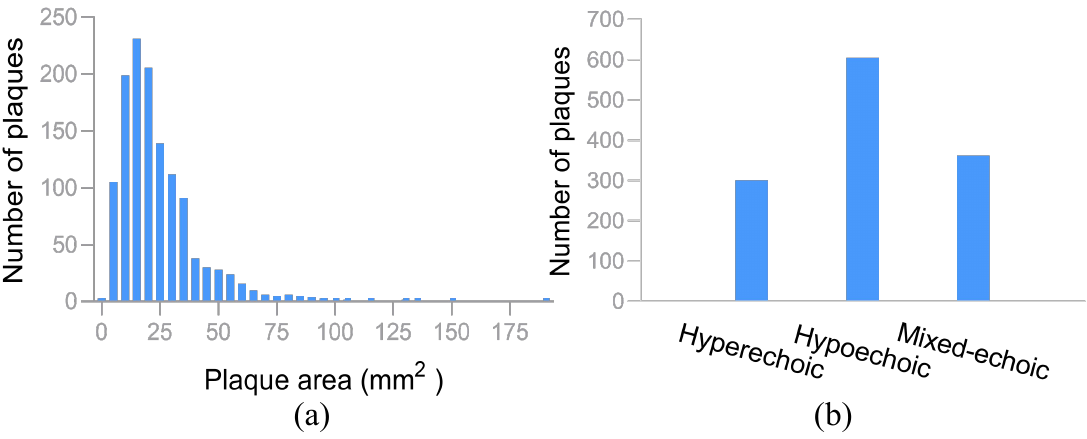}
	\end{center}
	\caption{Distribution of plaque area and plaque types for all 1270 ultrasound images. (a) Distribution for different plaque areas with an interval of 5 mm$^2$. (b) Distribution of there types of plaques (i.e., hyperechoic plaque, hypoechoic plaque, and mix-echoic plaque).}
	\label{fig:distribution}
\end{figure}

Figure \ref{fig:distribution} shows the plaque distribution of our datasets. The manual plaque area (PA) measurements have a mean of 24.03 mm$^2$ and range from 1.45 mm$^2$ to 187.06 mm$^2$ with a majority in the range of 10 mm$^2$ to 80 mm$^2$. The dataset contains 301 hyperechoic plaques, 605 hypoechoic plaques and 362 mixed-echoic plaques. 

Before algorithm implementation, we reduced the size of the examined 2D ultrasound images by cropping a region-of-interest around each plaque, as would be done during the clinical examination. The cropped images were automatically resized to 96$\times$144 matrices as network inputs. 

\subsection{Experiment Setting}
During the training of the multitask network, the dataset was randomly divided into training, validation, and test sets in the ratio of 6:2:2, resulting in 751, 258 and 261 images in the three sets, respectively. The following parameters were used for network training: the number of epochs=200, batch size=10, optimizer=ADAM, and learning rate=1e-3. The hyperparameters $\alpha_1$, $\alpha_2$, $\alpha_3$ and $\alpha_4$ in the CCM were set to 0.1, 0.2, 0.3 and 0.4, respectively. All deep learning networks were implemented using Pytorch 1.10.0, CUDA 11.6, and Python 3.8 platforms on an NVIDIA GTX3090 GPU. 

Ultimately, all experimental results in this paper were obtained from three randomized experiments by setting different random seeds in the three repetitions. The segmentation evaluation metrics, DSC, ASSD, HD and $|\Delta PA|$, are expressed as mean$\pm$SD for the mean measurements of all patients, and the classification metrics, ACC, Precision, F1-score and Kappa, are presented as mean$\pm$SD of results generated by the three randomized experiments.

\subsection{Experimental results}
\subsubsection{Segmentation Accuracy of the Multi-task Framework}

\begin{figure}[!htbp]
	\begin{center}
		\includegraphics[width=0.5\textwidth,height=\textheight,keepaspectratio]{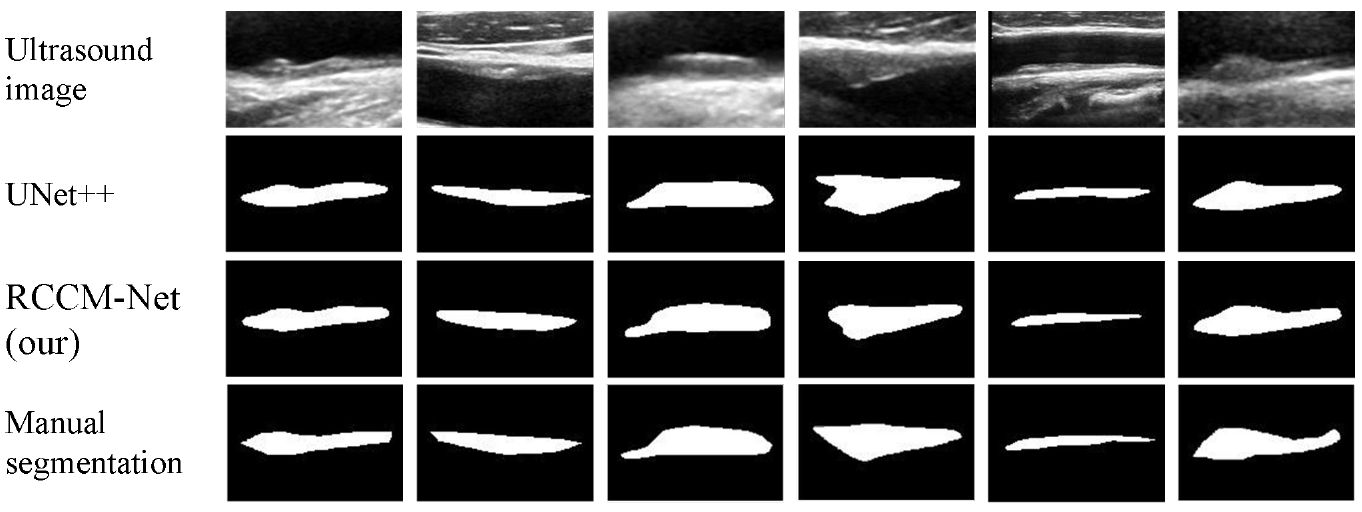}
	\end{center}
	\caption{Comparison of the carotid plaque segmentation results generated by our RCCM-Net and the baseline (UNet++). The first row shows the original ultrasound images of the plaques. Row 2 shows the UNet++-generated segmentations. Row 3 shows our multi-task algorithm-generated segmentations. The last row shows manual segmentations.}
	\label{fig:contours}
\end{figure}

Figure \ref{fig:contours} shows segmentations using the RCCM-Net approach compared to the baseline UNet++. The multi-task algorithm-generated segmentations outperformed the baseline UNet++ while generating results very close to the manual segmentations.  

\begin{figure}[!htbp]
	\begin{center}
		\includegraphics[width=0.5\textwidth,height=\textheight,keepaspectratio]{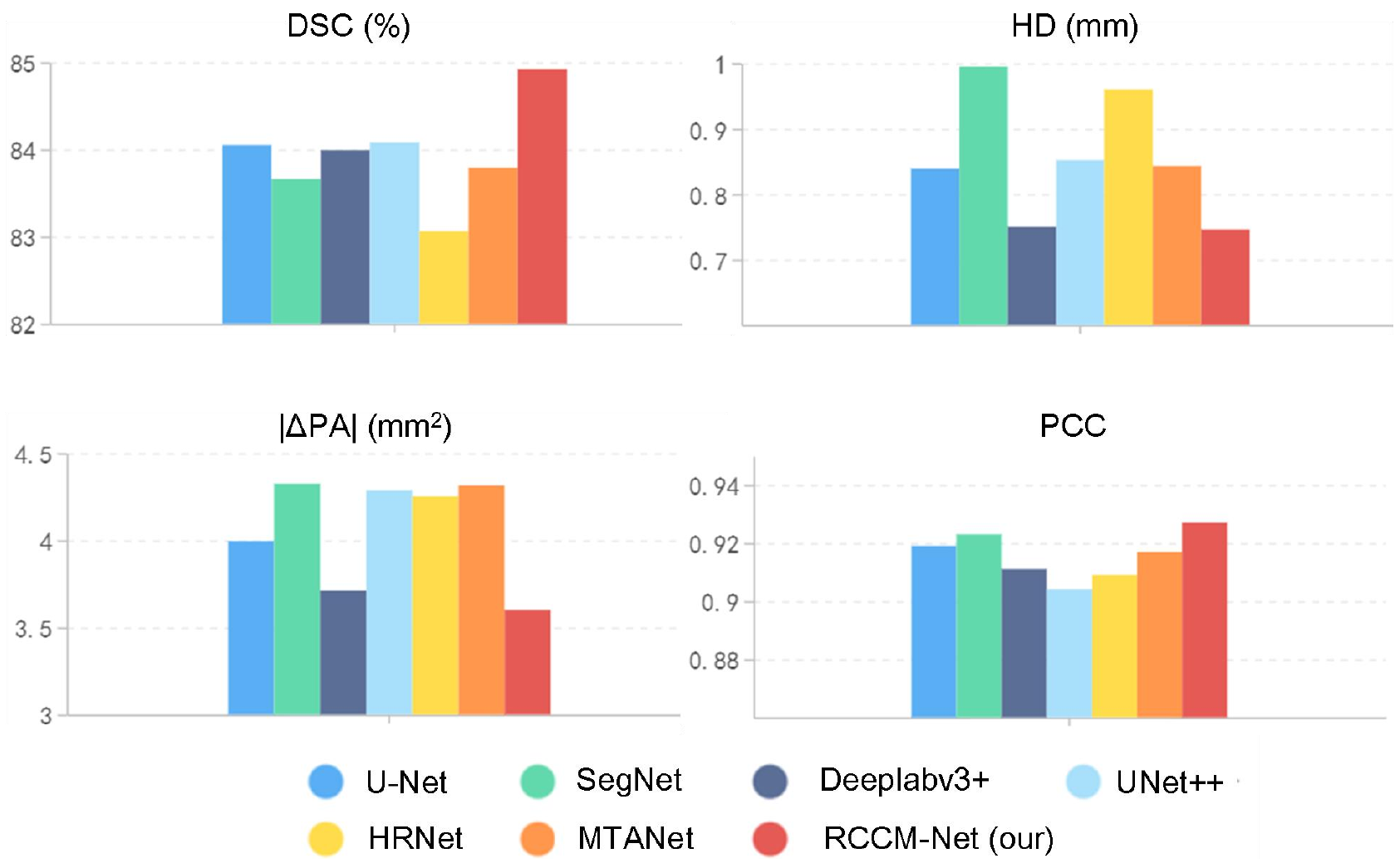}
	\end{center}
	\caption{Bar charts comparing RCCM-Net to existing single segmentation and multi-task networks in DSC, HD, $|\Delta PA|$ and PCC.}
	\label{fig:hist_seg}
\end{figure}

To evaluate the effectiveness of our proposed RCCM-Net, we further compared the performance of our algorithm to networks used for segmentation only, including U-Net \cite{zhou2021deep}, SegNet \cite{badrinarayanan2017segnet}, Deeplabv3+ \cite{chen2018encoder}, as well as to the baseline UNet++. Moreover, RCCM-Net is also compared to a state-of-art multi-task method HRNet \cite{wang2020deep} and MTANet \cite{ling2023mtanet}. Table \ref{tab:SegPerformance} and Figure \ref{fig:hist_seg} show the results of carotid plaque segmentation. RCCM-Net achieves better performance than the existing single segmentation networks (i.e., U-Net, SegNet, HRNet and UNet++) for all metrics. Compared to the baseline UNet++, RCCM-Net increased the DSC over the three different backbones by 1.15\%. reduced the ASSD, HD and $|\Delta PA|$ by 5.65\%, 11.19\% and 16.15\%, respectively, and increased the Pearson correlation coefficient from 0.889 to 0.927. Compared to the existing multi-task network (HRNet and MTANet), RCCM-Net also yielded better performance, especially in the improvements of distance metrics and plaque areas. 

\begin{table*}[!htbp]
	\centering
	\captionsetup{width=0.75\textwidth}{
	\caption{Comparison of the segmentation performance of our algorithm to different single segmentation and multi-task networks}
	\label{tab:SegPerformance}}
	\renewcommand\arraystretch{1.2}
	\begin{tabular}{|l|l|l|l|l|l|l|}
		\hline
		Method             & Type                         & DSC (\%)   & ASSD (mm)   & HD (mm)     & $|\Delta$PA$|$ (mm$^2$) & PCC         \\ \hline
		U-Net \cite{zhou2021deep}     & \multirow{4}{*}{Single task} & 84.05$\pm$6.76 & 0.289$\pm$0.173 & 0.839$\pm$0.693 & 3.995$\pm$5.543 & 0.919$\pm$0.011 \\
		SegNet \cite{badrinarayanan2017segnet}    &                              & 83.66$\pm$7.06 & 0.313$\pm$0.203 & 0.995$\pm$0.937 & 4.325$\pm$5.263 & 0.923$\pm$0.008 \\
		Deeplabv3+ \cite{chen2018encoder} &                              & 83.99$\pm$6.93 & 0.270$\pm$0.140 & 0.750$\pm$0.509 & 3.711$\pm$5.446 & 0.911$\pm$0.008 \\
		UNet++ \cite{zhou2019unet++}    &                              & 84.08$\pm$7.08 & 0.288$\pm$0.177 & 0.852$\pm$0.705 & 4.287$\pm$5.865 & 0.904$\pm$0.040 \\ \hline
		HRNet \cite{wang2020deep}       & \multirow{3}{*}{Multi-task}  & 83.06$\pm$6.91 & 0.310$\pm$0.205 & 0.960$\pm$0.854 & 4.252$\pm$5.408 & 0.909$\pm$0.030 \\
		MTANet  \cite{ling2023mtanet}   &                              & 83.79$\pm$7.05 & 0.285$\pm$0.193 & 0.843$\pm$0.840 & 4.314$\pm$5.848 & 0.917$\pm$0.008 \\
		RCCM-Net (our)     &                              & \textbf{84.92$\pm$7.16} & \textbf{0.267$\pm$0.164} & \textbf{0.746$\pm$0.543} & \textbf{3.599$\pm$5.299} & \textbf{0.928$\pm$0.016} \\ \hline
	\end{tabular}
    
\end{table*}

\begin{table*}[!htbp]
	\centering
	\captionsetup{width=0.65\textwidth}{
		\caption{Comparison of the classification performance of our algorithm to different single segmentation and multi-task networks.}
	\label{tab:ClsPerformance}}
	\renewcommand\arraystretch{1.2}		
	\begin{tabular}{|l|l|l|l|l|l|}
		\hline
		Method                & Type                         & ACC (\%)            & Precision (\%)      & F1-score             & Kappa                \\ \hline
		EfficientNet \cite{tan2019efficientnet} & \multirow{6}{*}{Single task} & 75.09$\pm$4.88          & 75.44$\pm$4.73          & 0.754$\pm$0.044          & 0.609$\pm$0.075          \\
		ResNet \cite{he2016deep}       &                              & 83.27$\pm$0.96          & 84.08$\pm$0.87          & 0.829$\pm$0.001          & 0.734$\pm$0.016          \\
		DenseNet \cite{huang2017densely}     &                              & 81.61$\pm$0.54          & 81.68$\pm$0.19          & 0.813$\pm$0.012          & 0.710$\pm$0.013          \\
		RepVGG \cite{ding2021repvgg}       &                              & 77.26$\pm$2.22          & 77.96$\pm$2.09          & 0.760$\pm$0.025          & 0.635$\pm$0.037          \\
		Res2Net \cite{gao2019res2net}       &                              & 81.86$\pm$2.77          & 82.24$\pm$2.67          & 0.818$\pm$0.028          & 0.714$\pm$0.044          \\
		DPN \cite{chen2017dual}          &                              & 79.44$\pm$2.46          & 79.16$\pm$3.26          & 0.790$\pm$0.028          & 0.676$\pm$0.037          \\ \hline
		HRNet \cite{wang2020deep}        & \multirow{3}{*}{Multi-task}  & 80.07$\pm$3.29          & 81.25$\pm$2.78          & 0.798$\pm$0.032          & 0.682$\pm$0.054          \\
		MTANet  \cite{ling2023mtanet}      &                              & 82.63$\pm$0.78          & 82.70$\pm$0.61          & 0.817$\pm$0.006          & 0.757$\pm$0.050          \\
		RCCM-Net (our)        &                              & \textbf{85.82$\pm$1.36} & \textbf{86.32$\pm$1.72} & \textbf{0.854$\pm$0.013} & \textbf{0.775$\pm$0.021} \\ \hline
	\end{tabular}
\end{table*}

\begin{figure*}[!htbp]
	\begin{center}
		\includegraphics[width=0.8\textwidth,height=\textheight,keepaspectratio]{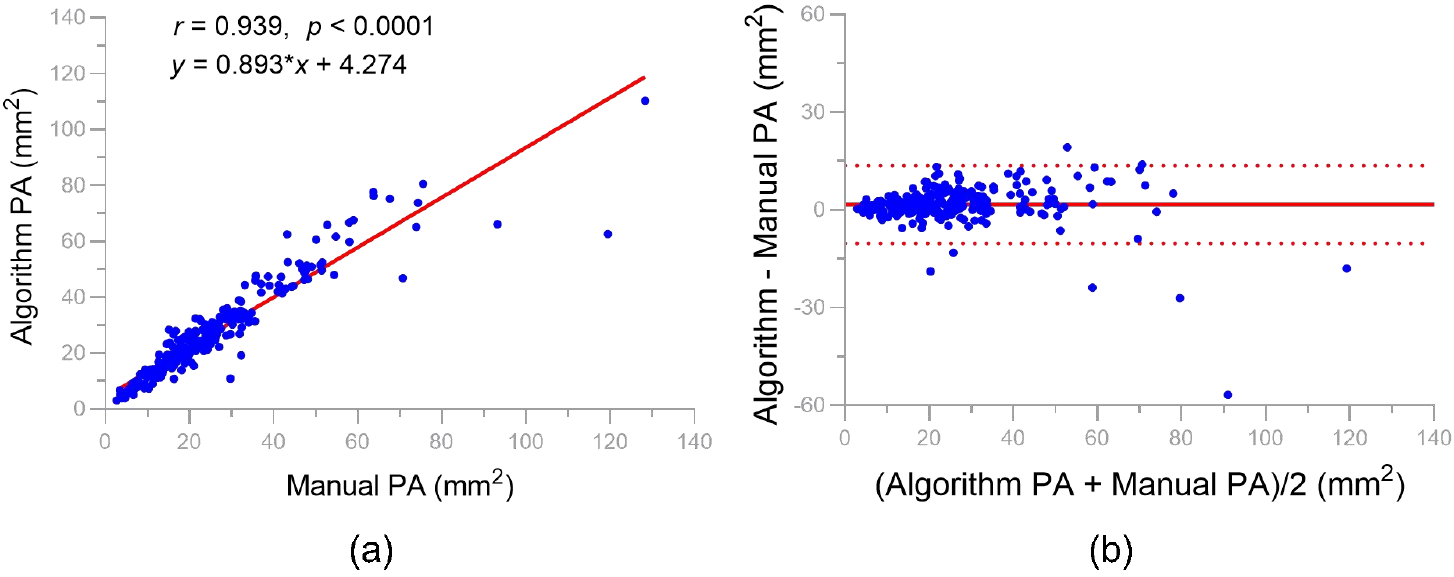}
	\end{center}
	\caption{Relationships of our RCCM-Net and manually generated plaque area (PA) measurements (n=261) from one of the random experiments. (a) Linear correlation ($r$=0.939, $p<0.0001$), and (b) Bland-Altman plot of the two sets of plaque area measurements. The solid red line and the dashed red lines represent the bias (1.684 mm$^2$) and mean$\pm$1.96 SD, respectively.}
	\label{fig:correlation}
\end{figure*}

Figure \ref{fig:correlation} shows the correlation and Bland-Altman plots of the results of one of the three random experiments, comparing the multi-task algorithm and the manually generated plaque areas for 261 carotid ultrasound images. The correlation is strong and significant with a correlation coefficient of 0.939 ($p<0.0001$) (Figure \ref{fig:correlation}(a)) and a small bias of 1.68 mm$^2$ with limits agreement from -10.23 to 13.59 mm$^2$ (Figure \ref{fig:correlation}(b)).

\subsubsection{Classification Accuracy of the Multi-task Framework}

To evaluate the classification accuracy of our proposed multi-task framework, we compared the performance of our algorithm to that of different single classification networks, including EfficientNet \cite{tan2019efficientnet}, ResNet \cite{he2016deep}, DenseNet \cite{huang2017densely}, RepVGG \cite{ding2021repvgg}, Res2Net \cite{gao2019res2net} and DPN \cite{chen2017dual}. Moreover, the RCCM-Net framework is also compared to a state-of-art multi-task method HRNet \cite{wang2020deep} and MTANet \cite{ling2023mtanet}. 

\begin{figure}[!htbp]
	\begin{center}
		\includegraphics[width=0.5\textwidth,height=\textheight,keepaspectratio]{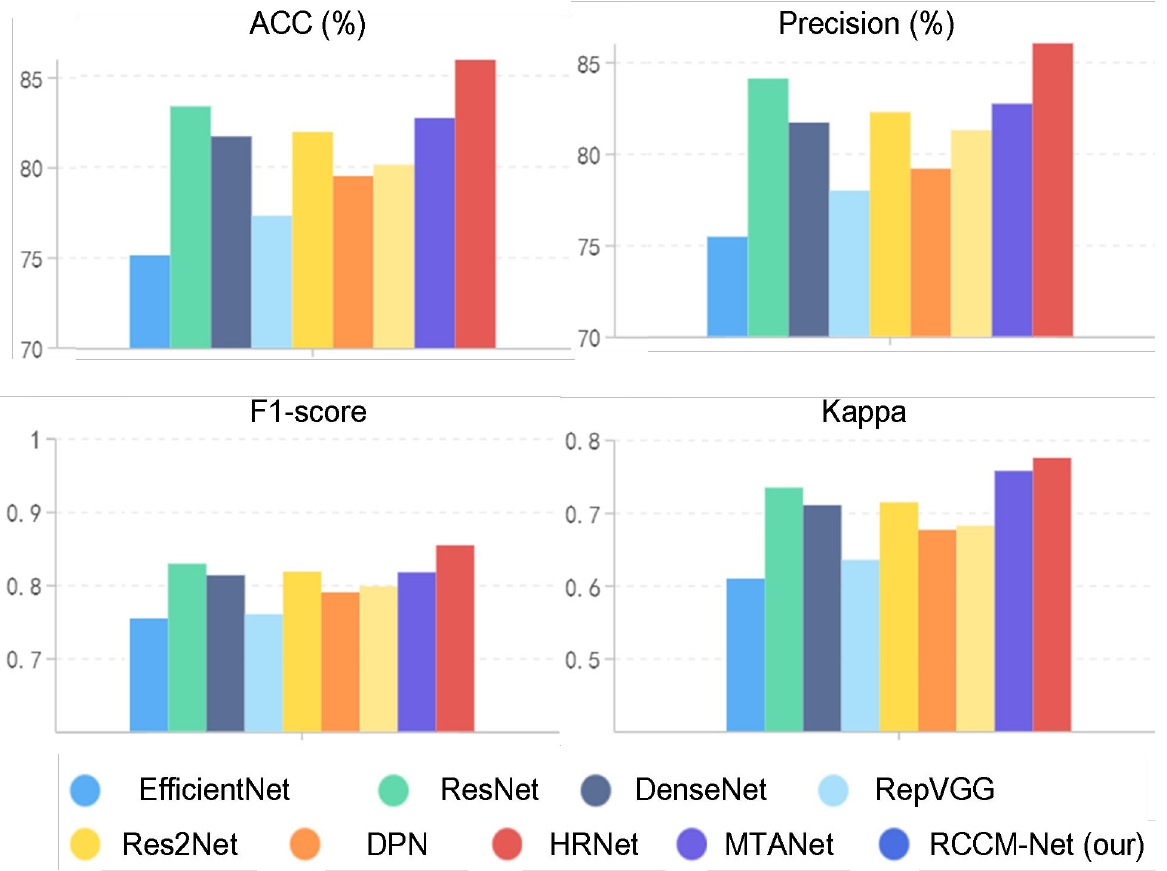}
	\end{center}
	\caption{Bar charts comparing RCCM-Net to existing single classification and multi-task networks in ACC, Precision, F1-score and Kappa coefficient.}
	\label{fig:hist_cls}
\end{figure}

Table \ref{tab:ClsPerformance} and Figure \ref{fig:hist_cls} summarize the carotid plaque classification performance of RCCM-Net and the results from the other networks. RCCM-Net achieved better performance than the existing single classification networks for all metrics. Moreover, use of RCCM-Net resulted in a greater ACC, Precision, F1-score and Kappa coefficient compared to the baseline (UNet++) by 3.06\%, 2.66\%, 3.02\% and 5.58\%, respectively. Compared to the existing multi-task network (HRNet and MTANet), RCCM-Net also yielded better performance, especially in greater accuracy and precision.

\begin{figure*}[!htbp]
	\begin{center}
		\includegraphics[width=0.8\textwidth,height=\textheight,keepaspectratio]{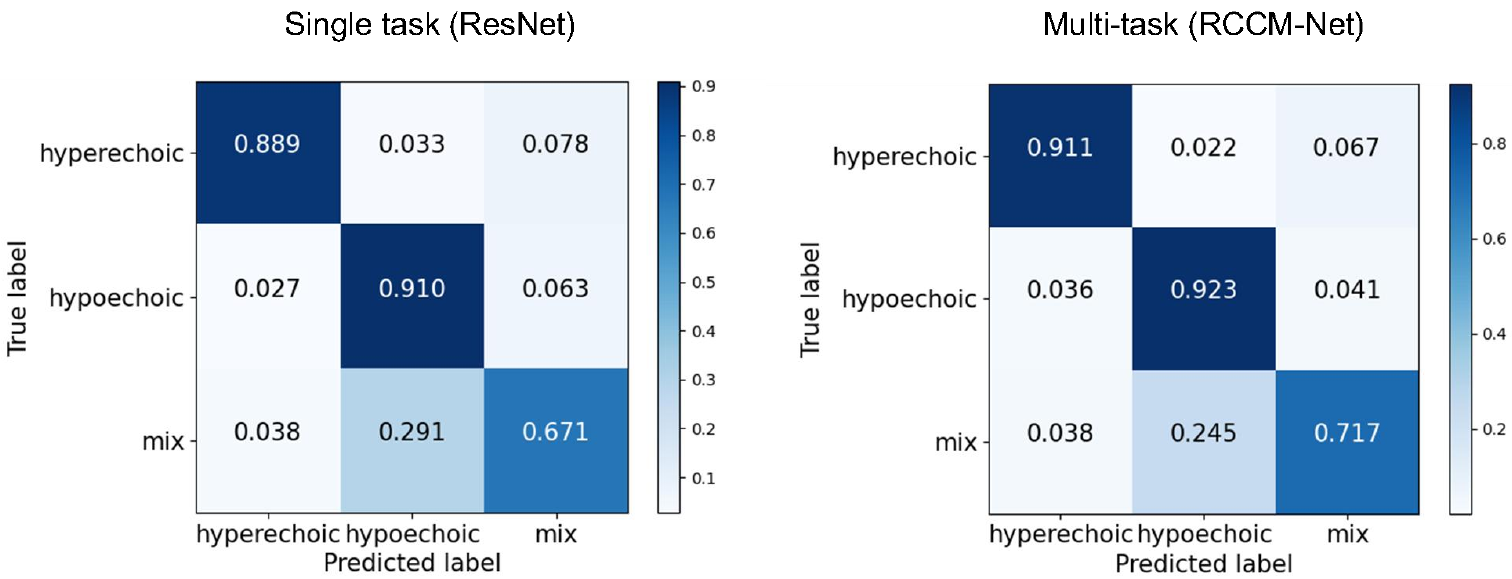}
	\end{center}
	\caption{Confusion matrix comparison of the single classification task with RCCM-Net.}
	\label{fig:confusion}
\end{figure*}

Figure \ref{fig:confusion} shows the carotid plaque classification performance as confusion matrices of the RCCM-Net model and the baseline UNet++. These results demonstrate that the multi-task approach improved the classification accuracy for all three networks. These results also show that the accuracy of classifying hyperechoic and hypoechoic plaques is higher than that of mixed-echoic plaques. 

\subsubsection{Ablation Experiments}
An ablation study was conducted to examine the effectiveness of the RCM and CCM modules. The baseline method (Base) is the base multi-task framework without the use of RCM and CCM. The experimental results are shown in Table \ref{tab:SegAblation} for plaque segmentation and in Table \ref{tab:ClsAblation} for plaque classification.

From Table \ref{tab:SegAblation}, we observe that using CCM enhanced the DSC, ASSD, and HD of the three backbones by 0.92\%, 1.43\%, and 7.37\%, respectively. $|\Delta PA|$ of Base+CCM is very close to the Base results. This suggests that CCM improved the segmentation task by allowing it to focus on misclassified samples with the help of the classification task. Furthermore, by adding the RCM module to the multi-task framework, we observe that Base+RCM also improved the Base results in all metrics, indicating that RCM can also improve the segmentation performance. Base+CCM+RCM achieves the best performance for the DSC, ASSD and HD metrics.

\begin{table}[]
	\centering
	\caption{Ablation experiment results for segmentation.}
	\label{tab:SegAblation}	
	\renewcommand\arraystretch{1.2}
	\resizebox{0.5\textwidth}{!}{
	\begin{tabular}{|p{0.4cm} | p{0.4cm} | p{0.5cm} | p{1.2cm} | p{1.3cm} |p{1.3cm}|p{1.3cm}|}
		\hline
		\multicolumn{3}{|l|}{Modules}                               & \multirow{2}{*}{\begin{tabular}[c]{@{}l@{}}DSC\\    (\%)\end{tabular}} & \multirow{2}{*}{\begin{tabular}[c]{@{}l@{}}ASSD\\    (mm)\end{tabular}} & \multirow{2}{*}{\begin{tabular}[c]{@{}l@{}}HD\\    (mm)\end{tabular}} & \multirow{2}{*}{\begin{tabular}[c]{@{}l@{}}$|\Delta PA|$\\   (mm$^2$)\end{tabular}} \\ \cline{1-3}
		\multicolumn{1}{|l|}{Base} & \multicolumn{1}{l|}{CCM} & RCM &                                                                           &                                                                            &                                                                          &                                                                              \\ \hline
		\multicolumn{1}{|l|}{\checkmark}    & \multicolumn{1}{l|}{}    &     & 83.65$\pm$7.17                                                                & 0.280$\pm$0.171                                                                & 0.868$\pm$0.709                                                              & 3.562$\pm$5.303                                                                  \\
		\multicolumn{1}{|l|}{}     & \multicolumn{1}{l|}{\checkmark}   &     & 84.42$\pm$7.47                                                                & 0.276$\pm$0.179                                                                & 0.804$\pm$0.639                                                              & 3.712$\pm$5.220                                                                  \\
		\multicolumn{1}{|l|}{}     & \multicolumn{1}{l|}{}    & \checkmark   & 83.89$\pm$7.47                                                                & 0.294$\pm$0.206                                                                & 0.900$\pm$0.852                                                              & \textbf{3.560$\pm$4.584}                                                         \\
		\multicolumn{1}{|l|}{\checkmark}    & \multicolumn{1}{l|}{\checkmark}   & \checkmark   & \textbf{84.92$\pm$7.16}                                                       & \textbf{0.267$\pm$0.164}                                                       & \textbf{0.746$\pm$0.543}                                                     & 3.599$\pm$5.299                                                                  \\ \hline
	\end{tabular}}
\end{table}

Table \ref{tab:ClsAblation} shows that Base+RCM outperformed Base in almost all metrics, except for the F1-score of ResNet. Specifically, the accuracy, precision, and Kappa are greater by 0.31\%, 0.51\% and 0.40\%, respectively. Base+RCM achieved the same F1-score as Base. These results indicate that the classification task can learn new features from the segmentation task, i.e., the segmentation task can facilitate the classification task. Base+CCM achieved similar results to Base in classification metrics, because CCM is designed for segmentation, which improved the segmentation metrics (as shown in Table \ref{tab:SegAblation}). Furthermore, by adding the CCM module, Base+CCM+RCM yielded the best performance in all metrics.

\begin{table}[]
	\centering
	\caption{Ablation experiment results for classification.}
	\label{tab:ClsAblation}	
	\renewcommand\arraystretch{1.2}
	\resizebox{0.5\textwidth}{!}{
	\begin{tabular}{|p{0.4cm} | p{0.4cm} | p{0.5cm} | p{1.2cm} | p{1.3cm} |p{1.3cm}|p{1.3cm}|}
		\hline
		\multicolumn{3}{|l|}{Modules}                               & \multirow{2}{*}{\begin{tabular}[c]{@{}l@{}}ACC\\ (\%)\end{tabular}} & \multirow{2}{*}{\begin{tabular}[c]{@{}l@{}}Precision\\ (\%)\end{tabular}} & \multirow{2}{*}{F1-score} & \multirow{2}{*}{Kappa} \\ \cline{1-3}
		\multicolumn{1}{|l|}{Base} & \multicolumn{1}{l|}{CCM} & RCM &                                                                     &                                                                           &                           &                        \\ \hline
		\multicolumn{1}{|l|}{\checkmark}    & \multicolumn{1}{l|}{}    &     & 84.29$\pm$0.94                                                          & 84.64$\pm$1.05                                                                & 0.839$\pm$0.010               & 0.751$\pm$0.015            \\
		\multicolumn{1}{|l|}{}     & \multicolumn{1}{l|}{\checkmark}   &     & 84.07$\pm$1.10                                                          & 84.64$\pm$1.19                                                                & 0.837$\pm$0.012               & 0.747$\pm$0.017            \\
		\multicolumn{1}{|l|}{}     & \multicolumn{1}{l|}{}    & \checkmark   & 84.55$\pm$1.41                                                          & 85.07$\pm$1.24                                                                & 0.839$\pm$0.018               & 0.754$\pm$0.025            \\
		\multicolumn{1}{|l|}{\checkmark}    & \multicolumn{1}{l|}{\checkmark}   & \checkmark   & \textbf{85.82$\pm$1.36}                                                 & \textbf{86.32$\pm$1.72}                                                       & \textbf{0.854$\pm$0.013}      & \textbf{0.775$\pm$0.021}   \\ \hline
	\end{tabular}}
\end{table}

\subsubsection{Computational Time}
The mean testing time for a single plaque segmentation and classification by the RCCM-Net framework was 32.09$\pm$4.15 ms per image, which is similar to a single segmentation or classification task. This time is sufficiently short to allow plaque analysis to be performed immediately after the carotid image acquisition.

\section{Discussion}
Accurate segmentation and classification of carotid plaques are crucial for identifying high-risk patients, selecting appropriate treatments, and monitoring the progression of atherosclerosis. In this study, we proposed a region and category confidence-based multi-task network (RCCM-Net) for carotid ultrasound image analysis that trains a single model to generate both a carotid plaque segmentation and plaque type simultaneously. Our experiments show a strong correlation between our algorithm and manual plaque areas (PA) and excellent accuracy in plaque classification using RCCM-Net. Furthermore, we show the mutually beneficial relationship between the segmentation and classification tasks in improving performance, as well as the efficiency of our algorithm to generate PA and plaque types simultaneously.

The proposed multi-task deep learning framework differs from the previous works that used training of a two-stage model and existing multi-task algorithms. In particular, the previous carotid ultrasound image analysis methods involved training separate segmentation and classification models and required two stages for testing  \cite{jain2021hybrid, saba2021multicenter, skandha20203}. However, these methods have several limitations, including segmentation errors leading to decreased classification accuracy and increased training time due to the need for two separate models. In contrast, our method provides a novel framework, aiming at using a single model to accomplish both carotid plaque segmentation and classification tasks. The previous multi-task algorithms did not consider the mutually beneficial relationship between the segmentation and classification tasks. For example, Shen et al. developed a multi-task learning method named NDDR-LCS that leveraged auxiliary information from ultrasound reports to assist the carotid plaque classification task \cite{shen2020nddr}. Ling et al. proposed a multi-task attention network (MTANet) by using the reverse addition attention to fuse areas in multi-level layers of UNet for medical image segmentation and classification \cite{ling2023mtanet}. The architecture of MTANet was more beneficial to the segmentation task from multi-level features and didn’t consider using the classification task to promote the segmentation performance. However, our RCCM-Net incorporated a region-weight module (RCM) and a sample-weight module (CCM) to exploit the relationship between these two tasks, enabling mutual promotion of the performance of the two tasks. 

The proposed multi-task algorithm was evaluated on a large dataset with 1270 longitudinal carotid ultrasound images collected in Zhongnan Hospital (Wuhan, China). The manual plaque area (PA) measurements had a mean of 24.03 mm$^2$ and range from 1.45 mm$^2$ to 187.06 mm$^2$ with a majority in the range of 10 mm$^2$ to 80 mm$^2$. The small areas of these plaques, representing early-stage disease, make accurate segmentation and classification difficult, while early identification and intervention in these patients can reduce the chance of stroke later in life. Nonetheless, the segmentation task of the developed multi-task algorithm yielded DSC, ASSD, HD and $|\Delta PA|$ in excellent agreement with the manual segmentation results. Compared to the widely used U-Net++ network, our approach achieved better performance by 1.15\%, 5.65\%, 11.19\% and 16.15\% for DSC, ASSD, HD and $|\Delta PA|$, respectively. The algorithm PA measurements were strongly and significantly ($r$=0.927 and $p<0.0001$) correlated with manual measurements. For classification, our algorithm also showed high accuracy and agreement. Compared to the popular classification networks (i.e., EfficientNet, ResNet, DenseNet, RepVGG, and HRNet), our approach achieved the highest performance for ACC, Precision, F1-score, and Kappa coefficient. These results indicate that our approach provided high performance for both plaque segmentation and classification tasks, possibly making our method possible to be used in clinical practice.

Furthermore, compared to the general multi-task algorithms, our method incorporated RCM and CCM to learn the relationship between plaque segmentation and classification tasks. The ablation results showed that RCM and CCM caused the two tasks to benefit from each other. For the segmentation results as shown in Table \ref{tab:SegPerformance}, the DSC, ASSD, HD, $|\Delta PA|$ improved with the help of CCM. For the classification results as shown in Table \ref{tab:ClsPerformance}, CCM also showed an increase in specificity, accuracy, precision, and Kappa. This is due to two reasons: (1) RCM integrates the probability map of segmentation results into the high-level features of the classification task, enabling the classification task to learn the location information of plaques; and (2) CCM utilizes prediction errors as sample weights in the loss function of the segmentation task, enabling the segmentation task to focus on training misclassified samples. These results suggest that the proposed RCM and CCM modules could increase the performance of the multi-task framework.

Although we achieved high algorithm segmentation and classification accuracy, we acknowledge some limitations. We note that the classification accuracy of mixed-echoic plaques is much lower than that of hyperechoic plaques and hypoechoic plaques. This might be because the complex structure of mixed-echoic plaques, which are more difficult to identify, may result in the misclassification of the annotations and classification. For future work, we will investigate more attention modules to make the networks focus on the features of plaque compositions. Furthermore, we will also implement this approach to other plaque burden measurements (i.e., Total plaque volume (TPV), Vessel wall volume (VWV)).

\section{Conclusion}
This work is the first to report on the development of a multi-task framework to generate carotid plaque area measurements and plaque types simultaneously, in which the region-weight module and the sample-weight module were developed to exploit the correlation between classification and segmentation. Experimental results on a dataset of early stage and small plaques show that the proposed approach can effectively improve the accuracy of carotid plaque classification and segmentation, suggesting it may be useful in clinical practice and clinical trials for monitoring patients being treated medically and for evaluating new therapies that may arrest the progression of atherosclerosis.

\section*{Acknowledgment}
The authors thank the Zhongnan Hospital research team for providing the 2D US images. We also thank the Liyuan Hospital research team for their efforts in manually identifying plaques from carotid ultrasound images.

\section*{Reference}
\bibliographystyle{IEEEtran}
\bibliography{Ref}
\end{document}